\documentclass[conference]{IEEEtran}
\usepackage{amsmath,amssymb,amsfonts,amsthm}
\usepackage{xcolor}
\usepackage{cite}
\usepackage{graphicx} 
\usepackage{algorithmic,algorithm}
\usepackage{extarrows}
\usepackage{comment}
\usepackage[acronym]{glossaries}
\usepackage{float}
\usepackage{caption}
\usepackage{array} 
\usepackage{makecell}
\usepackage{subcaption}

\def\BibTeX{{\rm B\kern-.05em{\sc i\kern-.025em b}\kern-.08em
    T\kern-.1667em\lower.7ex\hbox{E}\kern-.125emX}}

\ifCLASSINFOpdf

\else

\fi

\begin{document}

\title{A TRNG Implemented using a Soft-Data Based Sponge Function within a Unified Strong PUF Architecture}

\author{
    \IEEEauthorblockN{Rachel~Cazzola\IEEEauthorrefmark{1}, Cyrus~Minwalla\IEEEauthorrefmark{2}, Calvin Chan\IEEEauthorrefmark{3}, Jim~Plusquellic\IEEEauthorrefmark{1}}
    \IEEEauthorblockA{\IEEEauthorrefmark{1}Department of Electrical and Computer Engineering, University of New Mexico, New Mexico, USA}
    \IEEEauthorblockA{\IEEEauthorrefmark{2}Cloud and Automation Technologies, Bank of Canada, Ontario, Canada}
    \IEEEauthorblockA{\IEEEauthorrefmark{3}Department of Electrical, Computer and Energy Engineering, University of Colorado Boulder, Colorado, USA}
    \IEEEauthorblockA{E-mail: jimp@ece.unm.edu, minw@bank-banque-canada.ca, calvin.chan@colorado.edu}
}




\markboth{Transactions on Computer-Aided Design,~Vol.~x, No.~y, Month~Year}%
{Shell \MakeLowercase{\textit{et al.}}: Bare Demo of IEEEtran.cls for Journals}

\maketitle

\begin{abstract}%
Hardware security primitives including True Random Number Generators (TRNG) and Physical Unclonable Functions (PUFs) are central components to establishing a root of trust in microelectronic systems. In this paper, we propose a unified PUF-TRNG architecture that leverages a combination of the static entropy available in a strong PUF called the shift-register, reconvergent-fanout (SiRF) PUF, and the dynamic entropy associated with random noise present in path delay measurements. The SiRF PUF uses an engineered netlist containing a large number of paths as the source of static entropy, and a time-to-digital-converter (TDC) as a high-resolution, embedded instrument for measuring path delays, where measurement noise serves as the source of dynamic entropy. A novel data post-processing algorithm is proposed based on a modified duplex sponge construction. The sponge function operates on soft data, i.e., fixed point data values, to add entropy to the ensuing random bit sequences and to increase the bit generation rate. A post-processing algorithm for reproducing PUF-generated encryption keys is also used in the TRNG to protect against temperature-voltage attacks designed to subvert the random characteristics in the bit sequences. The unified PUF-TRNG architecture is implemented across multiple instances of a ZYBO Z7-10 FPGA board and extensively tested with NIST SP 800-22, NIST SP 800-90B, AIS-31, and DieHarder test suites. Results indicate a stable and robust TRNG design with excellent min-entropy and a moderate data rate. 
\end{abstract}



\IEEEpeerreviewmaketitle

\section{Introduction}\label{Section:Introduction}

\IEEEPARstart{H}{ardware} security plays an increasingly important role in microelectronic systems, particularly those used to implement IoT resource-constrained applications and those used in unsupervised environments with heightened vulnerability to invasive attacks. The term ``hardware security module'' (HSM) is now widely used by security companies as an encapsulation of hardware security and trust primitives. HSMs build iron-clad security functions on top of a foundational module that is capable of generating random bit sequences, which are used either as keys for encryption, hashing, and authentication algorithms, as nonces for randomizing execution and authentication messages, or as initialization vectors for encryption. 

Physical unclonable functions (PUFs) have emerged as hardware security primitives capable of serving as the root of trust within HSMs for key generation. The strong connection between PUFs and TRNGs has led to the proposal of several unified architectures. The primary distinction between PUFs and TRNGs is related to their sources of entropy. PUFs leverage a static source of entropy, i.e., baked-in manufacturing variations that are unique to each device and ideally remain stable throughout the lifecycle of the device. TRNGs target dynamic entropy, such as jitter, chaos, metastability, and sources of physical noise (e.g., thermal, shot, $1/f$). 

Another important distinction relates to components of the architecture that are responsible for ensuring reproducibility of the bitstring. TRNGs have no such requirements, so these components are typically not needed. However, since TRNGs are subject to temperature-voltage attacks, we propose to leverage the PUF's reliability-enhancing module to add resilience against such attacks. A third distinction is related to the number of bits required and the bit generation rate, where TRNGs need to out-pace PUFs by orders of magnitude. Last, unlike TRNGs, the static source of entropy leveraged by PUFs is fixed and limited, significantly reducing the size of the entropy pool available to a TRNG. Therefore, TRNGs must be based on an unlimited dynamic source of entropy, or a combination of static and dynamic entropy, to be capable of meeting the demands of the HSM.

On the other hand, TRNGs and PUFs also have common requirements, namely the ability to produce random and unique bitstrings. Therefore, data post-processing components that improve the randomness of bitstrings, or improve their uniqueness across devices, can be used by both security functions. The common requirements of PUFs and TRNGs make a unified PUF-TRNG architecture attractive because a unified architecture can be smaller in size when compared to the combined sizes of the stand-alone versions. Moreover, both security functions are required in HSMs to provide a full range of security services to a wide range of application environments.

A unified PUF-TRNG hardware security primitive capable of serving the aforementioned roles is proposed and demonstrated in this paper. The key generation capability of the shift-register, reconvergent-fanout (SiRF) strong PUF is leveraged here as a source of static entropy that is combined with a dynamic source of entropy to define a true random number generator (TRNG). The unified PUF-TRNG architecture re-uses the data measurement and post-processing functions implemented within the SiRF PUF algorithm, adding only one additional module. The size of the expanded architecture is only approximately 5\% larger. 

A novel modification to the PUF data post-processing algorithm is proposed, introducing a functionality akin to the sponge function used in modern hashing algorithms (ex. SHA-3). However, instead of operating on bitstrings, the proposed sponge construction performs permutations on a set of soft-data values, i.e., fixed point values in the range of $\pm64$. The soft-data values are propagated from one iteration to the next in a chaining fashion, where each iteration updates the soft-data values based on a sample of the path delay measurements and a set of randomized parameters. This chaining approach has the advantage of eliminating all correlations that occur when reusing the path delay measurements over multiple iterations. Moreover, using soft-data expands the capacity of the internal state, thereby increasing the size of the entropy pool.

The specific contributions of this work include:

\begin{itemize}
   \item A unified PUF-TRNG architecture that reuses more than 95\% of the functionality of the stand-alone PUF architecture. The architecture includes a mode switch that changes the behavior of the challenge generation and the data post-processing operations associated with the PUF and TRNG functions. 
   \item A novel soft-data based sponge construction that enables the use of both static and dynamic sources of entropy by the TRNG, while eliminating all traces of correlation from the static source.
   \item A Global Process and Environmental Variation (GPEV) post-processing module that adds resilience to temperature-voltage attacks.
   \item Experimental results validating the TRNG design and acceptance testing in four of the most popular random number test tools. 
\end{itemize}

The remainder of this paper is organized as follows. Section \ref{Section:RelatedWork} presents an overview of previously proposed unified PUF-TRNG architectures, and selected works describing stand-alone TRNG architectures. Section \ref{Section:SystemOverview} describes the SiRF PUF-TRNG architecture and algorithm.
Section \ref{Section:ExperimentalResults} presents the results of applying the statistical testing tools to the random bit sequences from multiple Xilinx Zynq SoC devices, and Section \ref{Section:Conclusions} presents our conclusions.

\section{Related Work} \label{Section:RelatedWork}

A combined PUF-TRNG design using an array of ring oscillators (ROs) to generate entropy through jitter measurements was introduced in \cite{Schaumont2009}. However, this architecture lacked mechanisms to counter temperature and voltage fluctuations and relies solely on noise as a source of entropy for the TRNG. Building on this foundation, the authors in \cite{MartinezGomez2020} developed a calibration method to enhance performance, yet the underlying RO structure remained susceptible to attacks involving machine learning. In a different approach, the authors in \cite{Larimian2020} presented a unified PUF-TRNG architecture built on embedded flash memory, exploiting both spatial and temporal current variations. Fabricated using Global Foundries’ 55 nm process, this design demonstrated resilience against machine learning-based intrusions. In more recent work, the authors in \cite{Taneja2021} implemented a unified PUF-TRNG architecture based on SRAM. Although this approach is scalable and offers high-speed operation, it had a significant LSB bit error rate, starting around 2.0\% and increasing to 4.8\% under temperature gradients. In contrast, the authors in \cite{ODonnel2004} proposed a hybrid PUF-TRNG architecture that addresses environmental variability by testing 1-bit path delay differences, but the algorithm lacks a predictable runtime, and no bit rate metrics are provided. The SiRF PUF presented in \cite{Plusquellic2022} also uses delay differences, but it contrastingly computes the delay across an extensive set of uniquely designed paths. Their approach draws from both fixed and random entropy sources and applies a distribution-based adjustment to counter environmental influences on multi-bit digitized delays. Until now, there have been no investigations of the SiRF PUF's ability to generate true random numbers. 

The authors of \cite{Cao2023} presented a compact, energy-efficient TRNG and PUF design for securing IoT devices, featuring a reconfigurable ring oscillator structure with on-chip calibration to ensure high entropy and reliability, and an authentication protocol to resist various attacks. Liu et al.~\cite{Liu2023} introduced a memristive TRNG with an intrinsic two-dimensional PUF for tamper-resistance in nanoelectronics, using unique physical entropy sources analyzed by a neural network to enhance security against cloning. While the two-dimensional PUF’s high sensitivity and randomness are beneficial for security, they might make the system more vulnerable to environmental variations (e.g., temperature or supply voltage changes), potentially impacting reliability. Pratihar et al.~\cite{Pratihar2021} presented a dual-mode PUF-TRNG design that leveraged both oscillation frequency and propagation delay, to provide both secure, instance-specific randomness for PUFs and high entropy for TRNGs. Although the design effectively integrated both functions with resistance to machine learning attacks, its reliance on specific Transition Effect Ring Oscillator (TERO) cell configurations may limit its adaptability to other circuit architectures or processes. The authors of \cite{Santiago2024} proposed a reconfigurable PUF-TRNG module achieving high hardware efficiency and robust cryptographic properties through ring oscillators and rigorous statistical testing. While the design improves hardware efficiency and scalability, the increased complexity of the configurable PUF-TRNG module may introduce potential challenges in terms of power consumption and processing overhead, especially in resource-constrained IoT environments. Satpathy et al.~\cite{Satpathy2019} designed a unified entropy generator combining a 512-bit CMOS entropy source with FPGA post-processing for secure IoT authentication. They achieved high throughput, low power, and resistance to power attacks, with 25\% area savings over separate PUF and TRNG. 

\begin{figure*}[t!]
    \centering
     \includegraphics[width=5.5in,keepaspectratio=true]{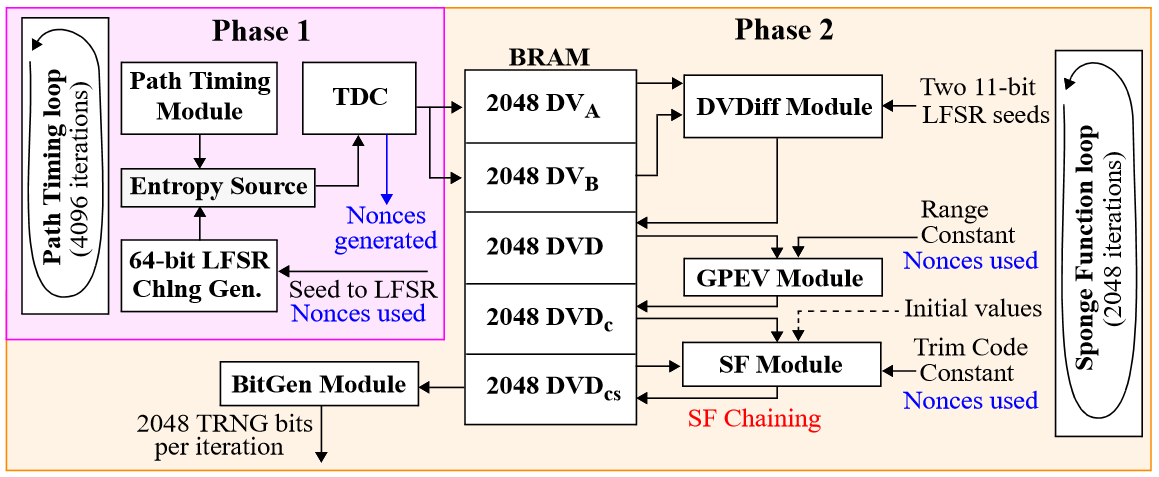}
     \caption{SiRF TRNG algorithm.}
     \label{Figure:SiRF_TRNG_algorithm}
     \vspace{-5pt}
\end{figure*}

In other work, Vatajelu et al.~\cite{Vatajelu2016} designed a unified PUF using magnetic RAM (MRAM) based on spin-transfer torque variations, which theoretically achieved zero-bit error rates. The PUF, however, was not physically realized, and its error rate remains untested in practice. Following this, Khan et al.~\cite{Khan2020} proposed a similar MRAM-based PUF and successfully fabricated and tested their model. Zalivako et al.~\cite{Zalivako2013} explored the use of a ring oscillator PUF as a source of randomness for generating true random numbers on an FPGA. Although the generated sequences demonstrated strong randomness, additional compression via an LFSR to improve statistical properties highlighted a limitation in the unprocessed PUF output. Sadr et al.~\cite{Sadr2012} presented a true random number generator using an Arbiter PUF within an NFSR, achieving 10 million bits per second with high entropy and low resource usage. The reliance on Arbiter PUFs however, may limit stability in environments with significant temperature or voltage fluctuations. The author of \cite{Cambou2018} presented a method for designing TRNGs using memory-based ternary PUFs, where unpredictable state cells generate multiple sources of randomness, which were enhanced by an XOR compiler or modulo-3 addition. Although the approach demonstrated high-quality randomness, the effectiveness of the XOR compiler is limited when the initial data stream lacks randomness. 

A stand-alone TRNG was proposed in \cite{Ingrid2022}, based on jitter noise within a ring oscillator based structure. The embedded carry chain within an Artix FPGA was used to obtain high-resolution measurements of three propagating edges launched simultaneously within the ring oscillator, which are processed into a random bit sequence. The authors run an extensive set of statistical tools to evaluate the quality of the random bit sequence. The authors of \cite{Chaos2020}, also propose a stand-alone TRNG that used a cellular automata topology to implement an asynchronous circuit structure capable of generating random bit sequences. The NIST SP800-90B and AIS-31 statistical test suites were applied in both publications, allowing for direct comparisons between their TRNGs and the proposed PUF-TRNG architecture described in this paper.

\vspace{26pt}
\section{System Overview} \label{Section:SystemOverview}

The proposed TRNG uses the SiRF PUF \cite{Plusquellic2022} static entropy source and data post-processing algorithms, as well as the dynamic entropy generated by the path delay measurement process. A flowchart of the operations carried out by the TRNG algorithm are shown in Fig. \ref{Figure:SiRF_TRNG_algorithm}. The algorithm can be partitioned into a random pattern generation, path timing, and nonce bit generation phase (Phase 1) and a data post-processing and random bit generation phase (Phase 2). Operations in Phase 1 are dedicated to measuring a set of path delays (static entropy) while simultaneously generating a set of nonce bits (dynamic entropy). 
These operations provide digitized timing data and nonce bits for randomizing several parameters utilized by the state machine modules of Phase 2, and for the LFSR in Phase I during subsequent iterations.

\subsection{Source of Static and Dynamic Entropy}

\begin{figure*}[t!]
    \centering
     \includegraphics[width=6.0in,keepaspectratio=true]{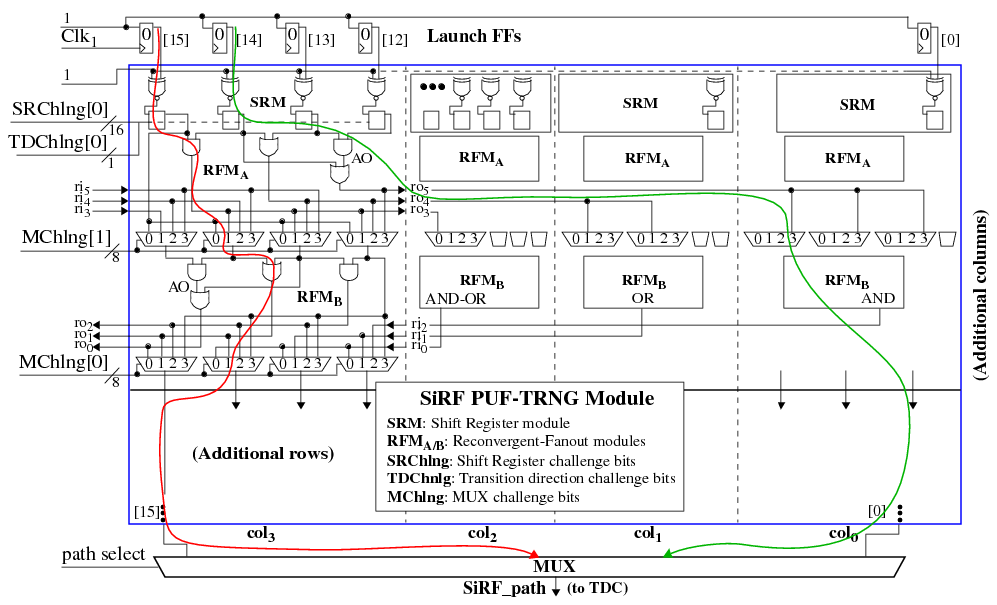}
     \caption{Basic building block of the SiRF PUF-TRNG architecture.}
     \label{Figure:SiRF_module}
     \vspace{-5pt}
\end{figure*}

The key building block of the static source of entropy for the SiRF PUF is shown in Fig. \ref{Figure:SiRF_module}, which consists of a sequence of shift-registers, non-inverting logic gates, and MUXs. The elements within the module's netlist are not fixed to specific layout positions, as is true for identically designed PUF architectures, and instead are placed and routed according to the optimization algorithms within the physical synthesis tool. We use wire constraints to prevent the place-and-route tool from modifying the netlist structure, and we design the netlist to ensure glitch-free propagation of signals. 

Challenges to the module, applied to the inputs along the left in the figure, are used to configure paths through the shift-registers and MUXs. The $TDChlng[0]$ bit of the challenge controls whether the rising transitions introduced by the \textbf{Launch FFs} on the module's inputs propagate through the module as rising (0) or falling (1) edges. This bit drives the low-order bit of the shift registers which determines the transition direction on the output of the shift registers. Since our design is based on the original SiRF netlist, additional details regarding its features are found in \cite{Plusquellic2022}.

The number of testable paths through each module with 16 primary outputs, as shown in Fig. \ref{Figure:SiRF_TRNG_algorithm}, is 512. The modules can be stacked vertically and horizontally. Each vertically stacked module multiplies the number of testable paths by a factor of 80. For a composite netlist with three rows and two columns ($3\times2$), the eight possible combinations of rising and falling transitions through each row result in 52,428,800 testable paths. The $3\times2$ configuration of modules is used as the netlist configuration in the experiments carried out in this work.

Dynamic entropy is represented as measurement noise in the path delay measurements, which is captured in the low-order bit of the delay value (DV). A full bit of dynamic entropy is obtained by XOR'ing the low-order bits of 12 consecutive path delay measurements. The need for 12 consecutive measurements was obtained from experiments carried out on FPGAs. We show in the Experimental Results section that this type of distillation process produces bitstrings that pass all statistical tests. During the path timing phase, a total of 341 nonce bits are obtained by measuring 4,096 path delays. 

The 341 nonce bits, or approximately 42 bytes, are used for randomizing two parameters in post-processing operations carried out in Phase II, as described below. The 2048 iterations of the Sponge Function loop shown in Fig. \ref{Figure:SiRF_TRNG_algorithm} reuse the 42 nonce bytes every 20 iterations of the loop, and therefore, the nonce bytes are reused at least 102 times over the 2048 iterations. As we show, the contributions by other sources of entropy in Phase II eliminate the penalty that can occur when sources of entropy are reused for multiple purposes.

\subsection{Phase 1 - Boot-Strap and DV Generation Operations}

A boot-strap operation labeled as Phase 1 in Fig. \ref{Figure:SiRF_TRNG_algorithm} is executed to obtain an initial set of nonce bits. Here, the initial seed to the 64-bit LFSR used in the \textit{Random Chlng Gen.} module to generate pseudo-random challenges is set to 1. The delay values (DV) measured during the boot-strap operation are then discarded, and only the nonce bits are retained.

The nonce bits are used to randomize several elements of the TRNG algorithm. The first 64 bits of the nonce are used to seed the 64-bit LFSR after the boot-strap operation to enable the \textit{Random Chlng Gen.} module to generate a second set of pseudo-random challenges. The Path Timing operation is started a second time to measure the delays of 4,096 paths. But unlike boot-strap, the DV are now stored in the BRAM.

The SiRF PUF-TRNG incorporates an embedded instrument referred to as the time-to-digital converter (TDC), which is used to measure path delays at a resolution of approximately 18 ps. The \textit{Random Chlng. Gen.} module generates a sequence of 128 random challenges using a 64-bit LFSR, where each challenge allows the measurement of exactly 32 path delays, one-at-a-time, for a total of 4,096 path delays. The TDC-digitized delays are referred to as delay values or DV, and can vary in value from approximately 300 to 1000 depending on the length of the path. The DV are stored as 12-bit integer values in an on-chip block RAM (BRAM) in two distinct groups of 2,048 elements each, labeled as $DV_A$ and $DV_B$ in Fig. \ref{Figure:SiRF_TRNG_algorithm}. Operations are carried out on these values in Phase 2.

\subsection{Phase 2 - Data Post-Processing and Random Bit Generation}

\begin{figure*}[t!]
    \centering
     \includegraphics[width=6.0in,keepaspectratio=true]{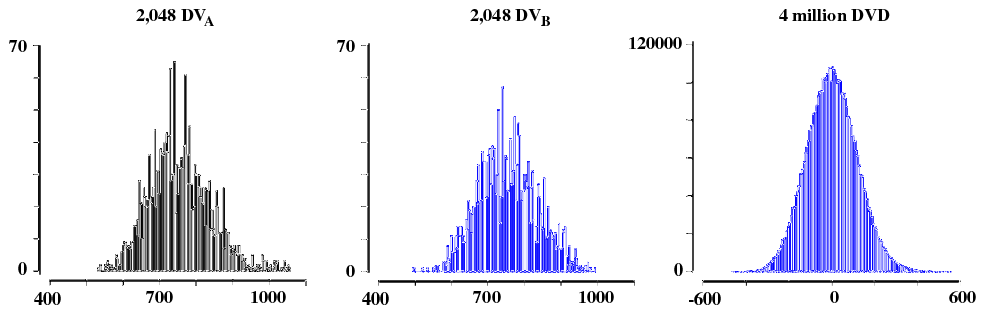}
     \caption{Example $DV_A$, $DV_B$ and $DVD$ path delay distributions. The $DVD$ are computed by creating differences using all combinations of elements in the $DV_A$ and $DV_B$ groups. }
     \label{Figure:DVA_DBV_DVD}
\end{figure*}

The SiRF algorithm consists of a sequence of four data post-processing modules labeled \textit{DVDiff}, \textit{GPEV}, \textit{SF} and \textit{BitGen}, as shown in Fig. \ref{Figure:SiRF_TRNG_algorithm}. The functions carried out by these modules are described in the following with illustrations to show the effect that they have on the DV. Each data post-processing module is executed 2,048 times, with each iteration producing 2,048 random bits. In tandem, these steps capture the necessary properties of a sponge function in a novel way. As shown on the right of \ref{Figure:SiRF_TRNG_algorithm}, we label the iteration of these post-processing modules as the \emph{Sponge Function loop}.

The sampled delay values $DV_A$ and $DV_B$ are combinatorially expanded into a set of $2048\times2048=2^{22}$ or 4 million digital value differences ($DVD$s). $DV_A$ and $DV_B$ are reused to accelerate the bit-generation speed. A full analysis of bit generation rate and resource utilization is provided in the Experimental Results section.


\subsubsection{Differencing}
The first module, \textit{DVDiff} module, uses two 11-bit LFSRs to pseudo-randomly select samples from the 2,048 $DV_A$ and $DV_B$ values. The 11-bit LFSRs are configured with a primitive polynomial enabling them to select all possible unique combinations of DV from the two sets. The seeds to the two LFSR at the beginning of each iteration are incremented and decremented, respectively, over the range from 0 to 2047 and from 2047 to 0, e.g., the first iteration assigns LFSR seeds 0 and 2047, the second iteration assigns 1 and 2046, etc. 
During each iteration, the selected elements in the $DV_A$ and $DV_B$ sets are pairwise subtracted to produce a set of 2,048 $DVD$, which are signed integers stored in the BRAM. A benefit of this approach is its simplicity, but, as we show, the drawback of full reuse is the existence of correlations between DVD sets. Example $DV_A$, $DV_B$ and $DVD$ distributions are shown in Fig. \ref{Figure:DVA_DBV_DVD}. 

\subsubsection{Temperature and Voltage Compensation}
Next, the Global Process and Environmental Variation (GPEV) step reads the 2,048 $DVD$ from the BRAM and applies two linear transformations. The first transformation removes delay variations introduced by both global process variations and temperature-supply voltage effects by effectively standardizing the $DVD$, while the second restores the $DVD$ distribution to an integer value range. The horizontal spread (range) associated with the second transformation is controlled by a randomized \textit{Range Constant} (RC) parameter. A 6-bit component of the nonces generated during boot-strap in Phase 1 is used to expand the range to values between 128 and 191, which adds unpredictability to the second transformation. The transformed DVD are stored in BRAM as $DVD_c$, where `c' refers to compensated. 

The first transformation is described in Eqs. \ref{Eq:ChipMean} through \ref{Eq:ChipRange}. It defends against temperature attacks, and supply voltage attacks that change the DC level. Voltage glitching that is applied during the path delay measurement process will disrupt the compensation process, but the impact that the attack has on the $DVD_c$ is unpredictable, and may in fact be defeated given the outlier avoidance method used in GPEV. The mean, $\mu$, of the DVD, is computed in the standard fashion. The range is computed as the width of the distribution using Eq. \ref{Eq:ChipRange}, with offsets defined by Eqs. \ref{Eq:MaxBound} and \ref{Eq:MinBound}. Here, the range is measured at the limits given by -5\% and +5\% of the maximum and minimum values, respectively, of elements in the distribution. The offsets make the range measurement robust to the presence of outliers. The parameter $rand\_RC$ in Eq \ref{Eq:DVDc} refers to the randomized Range Constant from Fig. \ref{Figure:SiRF_TRNG_algorithm}.

\begin{equation}
    \mu = \frac{\sum\limits_{j=1}^{|\text{DVD}|}\text{DVD}_j}{|\text{DVD}|}\label{Eq:ChipMean}\\
\end{equation}
\begin{equation}
    \text{max} = \max(\text{DVD}) -     0.05*\max(\text{DVD})\label{Eq:MaxBound}\\
\end{equation}
\begin{equation}
    \text{min} = \min(\text{DVD}) + 0.05*\min(\text{DVD})\label{Eq:MinBound}\\
\end{equation}
\begin{equation}
    \text{range} = \max_{\forall j \in |\text{DVD}|} \text{DVD}_j - \min_{\forall j \in |\text{DVD}|} \text{DVD}_j \label{Eq:ChipRange}\\
\end{equation}
\begin{equation}
    DVD_N = \frac{(\text{DVD} - \mu)}{\text{range}} \label{Eq:Nc}\\
\end{equation}
\begin{equation}
    \text{DVD}_c = DVD_N \times rand\_RC \label{Eq:DVDc}
\end{equation}




\subsubsection{Spread-Factor (SF) Chaining}
The Spread-Factor module, labeled \textit{SF} in Fig. \ref{Figure:SiRF_TRNG_algorithm}, is responsible for removing correlations, which occur when the same $DV_A$ and $DV_B$ are subtracted under all combinations by the DVDiff module over the 2,048 iterations of the Sponge Function loop. Spread-Factors (SF) refer to sets of digital values defined over the range given by Eq. \ref{Eq:SF}, that are subtracted from the $DVD_c$, to produce $DVD_{cs}$. The SF are also updated and re-used over consecutive iterations of the Sponge Function loop, as described below.


\begin{equation}
\begin{aligned}
    SF := \pm 64
\end{aligned}
    \label{Eq:SF}
\end{equation}

\begin{figure}[t!]
   \centering   \includegraphics[width=3.3in,keepaspectratio=true]{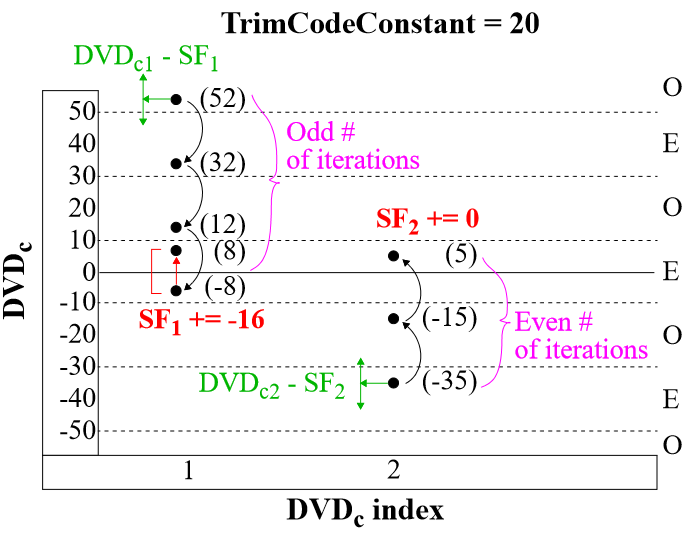}
   \caption{The randomized Trim Code Constant (TCC) parameter partitions the $DVD_c$ range into intervals shown by the dotted lines. The SF module processing operations are illustrated using two example $DVD_c$. }
   \label{Figure:SF_algorithm}
   \vspace{-5pt}
\end{figure}

The SF module defines a sequence of operations that are illustrated in Fig. \ref{Figure:SF_algorithm} using two $DVD_c$. The SF module accepts an input parameter called the \textit{Trim Code Constant} or TCC, whose value is randomized using a 3-bit nonce from the boot-strap operation in Phase 1. The 3-bit nonce is used to select an even-valued TCC between 8 and 22. The figure shows the operations carried out when the TCC selected is 20. 


The algorithm works as follows: First, the incoming $SF_x$ value is subtracted from the $DVD_{cx}$ value. Second, the SF module iteratively adds or subtracts the TCC from the $DVD_c$ until it falls with the region $\pm TCC/2$. And third, the number of subtractions or additions is used to determine if the current states of the $DVD_{cx}$ and $SF_x$ are updated. If the original $DVD_c$ is located in the odd 'O' region (annotated on the right side of the figure), an offset is computed that moves the $DVD_c$ to the symmetric position on the opposite side of the 0 line, and the offset is added to the $SF_x$. This occurs for $DVD_{c1}$ in Fig. \ref{Figure:SF_algorithm}, where the computed offset is -16. Otherwise neither the $DVD_{cx}$ nor $SF_x$ are changed from their current values, as shown for the $DVD_{c2}$ example in the figure.

The SF are set to zero on the first iteration of the Sponge Function loop (initialization is shown, labeled as \textit{Initial values} in Fig. \ref{Figure:SiRF_TRNG_algorithm}) and therefore, they have no effect on the $DVD_{cx}$ processed during the first iteration. For successive iterations, the $SF_x$ are randomly updated based on the selection of the TCC parameter and on the magnitude of the compensated difference values represented by the $DVD_{cx}$. The high order bit(s) of the $SF_x$ are modified as needed to maintain the $SF_x$ in the range of $\pm 64$ as a means of bounding their minimum and maximum values. 

\begin{figure}[H]
    \includegraphics[width=3.1in,keepaspectratio=true]{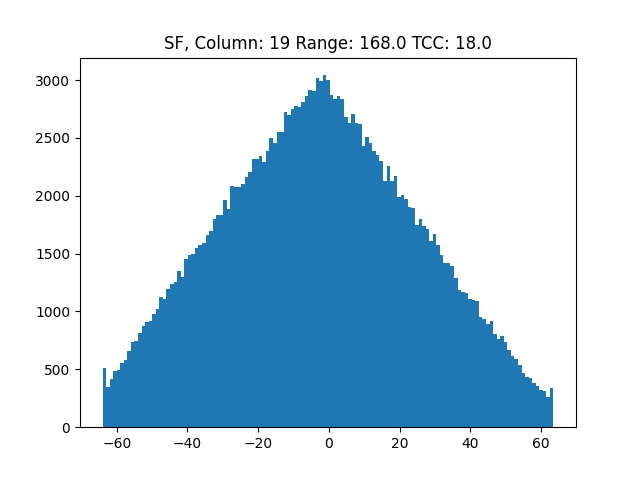}
    \caption{ Distribution of $SF$ generated during multiple iterations of the TRNG algorithm, starting at iteration 19, and then every 20th iteration thereafter where the Range is 168.0 and the TCC is 18.0. }
    \label{Figure:SF_distribution}
\end{figure}

The distributions of the $SF$ over successive iterations of the Sponge Function loop illustrate an important characteristic of the SiRF TRNG algorithm. The triangular shape of the distribution, shown in Fig. \ref{Figure:SF_distribution}, are typical of random distributions that are constructed by adding two discrete random numbers, e.g., the sums produced in experiments with two random die. The $SF$ in the current iteration are the sum of the incoming $DVD_c$ and the $SF$ from the previous iteration. The graph plots the set of 208,896 $SF$ produced over 102 iterations of the Sponge Function loop under the condition that the \textit{Range Constant} and \textit{Trim Code Constant} are the same (each iteration produces 2,048 $SF$, and the same RC and TCC are used every 20 iterations because of the nonce reuse discussed earlier). 
Fig. \ref{Figure:SF_distribution} is created by starting with iteration 19 of the Sponge Function loop. However, the other 19 distributions that can be created in this fashion are very similar.

\begin{figure}[H]
    \includegraphics[width=3.2in,keepaspectratio=true]{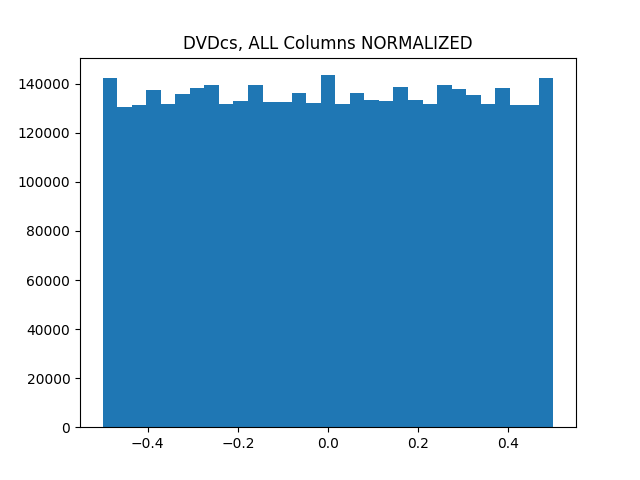}
   \caption{ Distribution of the $2^{22}$ \textit{normalized} $DVD_{cs}$ generated over all 2,048 iterations of the Sponge Function loop.}
   \label{Figure:DVDcs_distributions}
   \vspace{-5pt}
\end{figure}

In contrast, the $DVD_{cs}$ distributions generated as output in each iteration (and subsequently used to generate the random stream of bits) are uniformly distributed as shown by Fig. \ref{Figure:DVDcs_distributions}. Here, we divided each set of 2,048 $DVD_{cs}$ by the TCC used during that iteration, i.e., they are \textit{normalized} to values in the range between -0.5 and 0.5, to allow the underlying distribution for all $2^{22}$ $DVD{cs}$ to be illustrated.
Our analysis reveals that the number of negative and positive elements are nearly balanced, with a difference of only 812 elements or 0.02\%, across all $2^{22}$ $DVD_{cs}$. These two features of the TRNG function strongly support random statistical behavior, and explain the high statistical quality presented for the SiRF TRNG algorithm in the experimental results section.

\subsubsection{Bit Generation}
Until now, the DVDiff, GPEV, and SF modules absorbed and transformed fixed point values. The final step of our Sponge Function loop, BitGen, squeezes bits from these values, i.e., the $DVD_{cs}$ are processed into a sequence of 2,048 random bits by the \textit{BitGen} module. The \textit{BitGen} module generates a bit value of 0 if the $DVD_{cs}$ is negative, a bit value of 1 if the $DVD_{cs}$ is positive, and alternatives between generating a 0 and 1 in cases where the $DVD_{cs}$ is 0.0. The Sponge Function loop in Phase 2 is repeated for 2,048 iterations before Phase 1 operations are carried out to measure a new set of $DV_A$ and $DV_B$. 



\begin{figure}[H]
    \centering     \includegraphics[width=3.3in,keepaspectratio=true]{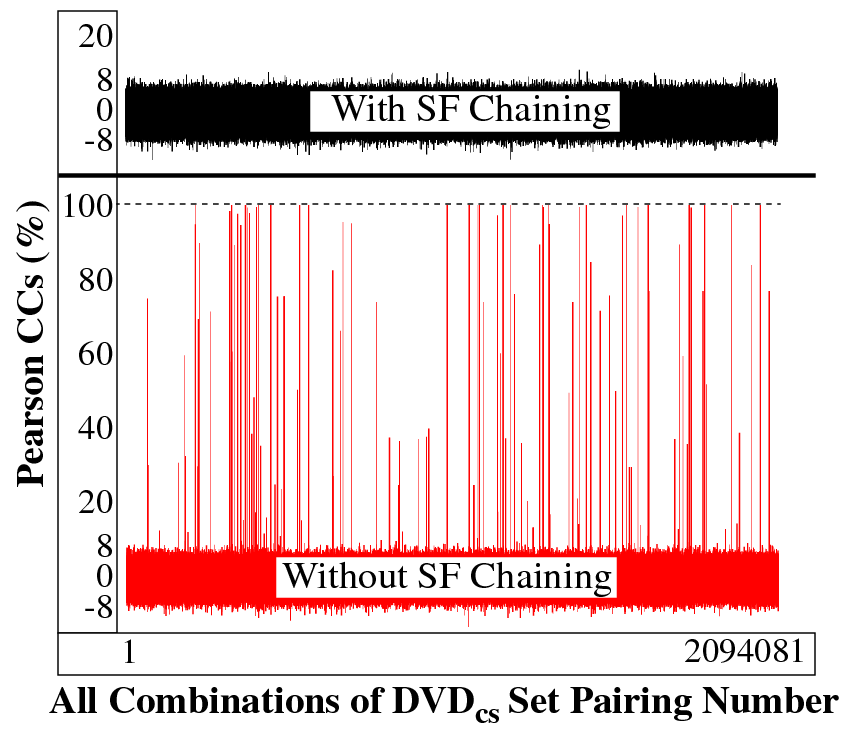}
     \caption{Pearson's correlation coefficients computed using all combinations of 2,048 element sets of $DVD_{cs}$ generated by Device $C_1$ over 2,048 iterations of the Sponge Function loop. The PCCs with SF Chaining are shown above those which do not use SF at all. }
     \label{Fig:PCCsAllIterations}
     \vspace{-5pt}
\end{figure}

\subsection{Analysis of Correlation With and without SF Chaining}

The SF Chaining operation is critical to achieving high statistical quality in the TRNG bit sequences. We ran NIST and DieHarder statistical test suites on the bit sequences generated with and without SF Chaining enabled, and only the bit sequences with SF chaining passed the tests. The statistical test results with SF Chaining are provided in Section \ref{Section:ExperimentalResults}. 

To better understand why SF Chaining is beneficial to the statistical quality of the bit sequences, we use a standard correlation test metric. The Pearson's Correlation Coefficient (PCC) measures the degree of similarity between two waveforms, and expresses that level over a range between -100\% to +100\%, where 0\% represents no correlation. We use PCC to determine if correlations exist in the 2,048-element $DVD_{cs}$ data sets. In order to comprehensively evaluate all possible sources of correlations, we compute PCCs using all pairing combinations of the $DVD_{cs}$ data sets. The PCCs are plotted in Fig. \ref{Fig:PCCsAllIterations}, where it is clear that there are many instances of maximum correlation at 100\% without SF Chaining (bottom). In contrast, with SF Chaining enabled (top), no pairing exceeds the value of $\pm$ 10\%, which indicates that SF Chaining is effective at removing all correlations that occur when the $DV_A$ and $DV_B$ are reused under all combinations by the DVDiff module.

The cases of 100\% correlation occur because some of the pairing sequences controlled by the two 11-bit LFSRs produce DVD sequences that are vertically shifted copies of each other. An intuitive example is fabricated as follows. Assume one DV from the $DV_A$ set is selected, and then the DVD are created by subtracting all of the elements of $DVD_B$ from this $DV_A$ element. Also, assume a second $DVD$ sequence is created in the same fashion but using a different $DV_A$ element. The shapes of two $DVD$ curves are identical and are only different by a vertical offset equal to the difference in the two $DV_A$ set elements. After GPEV is applied, the shift (DC offset) is removed, making the two curves identical, and 100\% correlated. Although this fabricated example is not possible when using two LFSRs to select elements, the LFSR pairing algorithm proposed cannot guarantee that all possible differences are generated and therefore, identical but shifted DVD sequences occur.

\subsection{Sponge Construction}

The post-processing steps that are chosen and implemented are deliberate in modeling the behavior of a cryptographic sponge construction \cite{Bertoni2010}. Specifically, the proposed approach is a random-permutation sponge based on the iterative nature of SF-chaining. An ideal sponge construction consists of absorption and squeezing phases, where absorption consists of one or more bit-wise operations while squeezing is a pseudo-random bit-space transformation. The two phases can be interleaved across subsets of the bitstring in a \textit{duplex} operating mode. It can be shown that the proposed TRNG achieves both absorption and squeezing properties via careful selection and ordering of the post-processing steps.  

The first step which calculates the $DV_{Diff}$ differences pseudo-randomly selects pairs of $DV$ elements from the available sets in the BRAM. While this is a soft-data operation, we can treat them as analogous to one or more bitwise operations (AND, XOR, etc.) applied in series. Note that bitwise operators do not preclude the presence of collisions, as witnessed in Fig. \ref{Fig:PCCsAllIterations}. The GPEV step is an additional absorption step that corrects for temperature and voltage variations. The SF Chaining step is a pseudo-random transformation that consumes values generated during the absorption phase. It is initialized with the original $DVD_c$ values that are subsequently permuted over multiple iterations. In each cycle, the function transforms each DVD by shifting it an arbitrary amount based on a randomly varying offset. The number of iterations, which is determined by the high-order bits of the SF-shifted $DVD_{cx}$, cannot be predicted \textit{a priori} because of the randomized TCC parameter. This crucial property ensures that the behavior of SF chaining over multiple iterations is equivalent to a pseudo-random permutation over the bit-space of $DVD_{cx}$. This process completely exhausts the underlying entropy of the stored delay values, while maximizing the throughput of the random bit generation process.

\section{Experimental Results} \label{Section:ExperimentalResults}

In this section, we present a statistical analysis using commonly used statistical tests, including NIST SP 800-22 \cite{NIST22}, NIST SP 800-90B \cite{NIST90B}, AIS 31 \cite{AIS31}, and DieHarder \cite{DieHarder} test suites. The data analyzed is collected from a set of five Zynq 7010 SoCs, installed on Digilent ZYBO boards \cite{ZYBO}. For the NIST SP 800-22 tests, we ran the TRNG repeatedly until each board generated 40 one-million bit (40 Mbit) sequences. For the NIST SP 800-90B and AIS 31 tests, we collected 10 MByte sequences from the five boards, while for Dieharder, the amount of data is unknown but in the range of 250 GigaBytes.

\subsection{NIST SP 800-22 Statistical Test Results}

The NIST SP 800-22 test suite consists of 15 distinct tests that measure, e.g., the frequency of 0's and 1's across the entire bitstring and locally over blocks of bits in the bit sequence, the length of the runs of 0's and 1's, the longest runs of 0's and 1's, and others referred to as cumulative sums, serial, compression, approximate entropy, etc. The 40 one-million bit TRNG bit sequences collected from each of the five Zynq devices pass all 15 tests, including all but one of the p-value-of-the-p-value tests. There was one instance of a failed non-overlapping template test, where only 36 of the 40 bitstrings passed, which is one less than the required number of 37.

\subsection{NIST SP 800-90B Statistical Test Results}

The NIST SP 800-90B test suite is a more recent addition that determines the pass-fail status of the random bit sequence. The 90B test suite evaluates the statistical quality of TRNG bit sequences using a conservative measure of entropy called \textbf{min-entropy}, which is traditionally defined as the amount of uncertainty in predicting the most-likely outcome from an entropy source. This NIST test suite estimates min-entropy using tests from two different tracks, the IID-track and the non-IID track, where TRNGs that sample from an Independent and Identically Distributed (IID) distribution employ the IID metrics. To determine the track, NIST recommends applying a set of tests to the TRNG bit sequence to determine if evidence can be found that the samples are not IID. If no evidence is found, then IID can be assumed.

The IID tests are pass-fail, and use a methodology called Permutation Testing. Permutation testing tests a statistical hypothesis in which the test statistic computed on a permuted version of the TRNG bit sequence is compared to the initial (un-permuted) sequence. The assumption is that permuting the initial bit sequence should produce similar test statistics. If \textit{t} represents the value of the test statistic on the original (un-permuted) sequence and \textit{t'} represents the test statistic computed on a permuted version of the original bit sequence, the IID tests count the number of times, over 10,000 permutations, that the value of \textit{t'} is less than \textit{t} (represented as $C_0$), and the number of times they are equal (represented as $C_1$). The IID test fails if the sum $C_0 + C_1 <= 5$ or if $C_0 >= 9,995$, indicating a significant difference exists between the permuted sequences and the original sequence. NIST repeats this evaluation using eleven IID statistical tests, listed and briefly explained in Table \ref{90B_IID_Explanation_Table}. A bit sequence passes the IID tests if and only if all eleven tests pass the count metric, otherwise the bit sequence is deemed non-IID. We subjected bit sequences of length of 10 MBytes for each of the five Zynq devices to the IID test suite and determined that all bit sequences pass the IID tests. 

\begin{table}[t!]
    \caption{NIST SP 800-90B IID Test Details}  
    \resizebox{\columnwidth}{!}{
    \vspace{-2.5pt}
    \centering 
    \renewcommand{\arraystretch}{1.7}
    \begin{tabular}{l | c }
    \hline
    \hline
    \makecell{\textbf{\normalsize{Test Name}}} & \bf \normalsize{Evaluation Criteria}  \\
    \hline
    \hline  
    \makecell{Excursion} & \makecell{Measures how far the running sum of sample values deviates \\ from its average value at each point in the bit sequence} \\
    \hline
    \makecell{Number of \\Directional Runs} & \makecell{Counts the number of runs of 0s and 1s across\\ consecutive samples} \\
    \hline    
    \makecell{Length of \\Directional Runs} & \makecell{Computes the length of the longest run across\\ consecutive samples} \\
    \hline
    \makecell{Number of Increases \\and Decreases} & \makecell{Counts the maximum number of increases or decreases\\ between consecutive sample values} \\
    \hline
    \makecell{Number of Runs\\ Based on the Median} & \makecell{Counts the number of runs that are constructed with \\respect to the median of the input data} \\
    \hline    
    \makecell{Length of Runs \\Based on Median} & \makecell{Determines the length of the longest run that is\\ constructed with respect to the median of the input data} \\
    \hline
    \makecell{Average Collision} & \makecell{Counts the number of successive sample values until \\a duplicate is found} \\
    \hline
    \makecell{Maximum Collision} & \makecell{Counts the maximum number of successive sample values\\ until a duplicate is found} \\
    \hline    
    \makecell{Periodicity} & \makecell{Determine the number of periodic structures in the data} \\
    \hline
    \makecell{Covariance} & \makecell{Measures the strength of the lagged correlation}\\
    \hline
    \makecell{Compression} & \makecell{Length of the compressed sequence (bit sequence\\ is first encoded)}\\
    \hline     
    \hline    
    \end{tabular}}
    \label{90B_IID_Explanation_Table}
    \vspace{-.5pt}
\end{table}

\begin{table}[t!]
    \centering 
    \caption{IID and Non-IID Test Results of the NIST SP 800-90B Entropy Estimation} 
    \resizebox{\columnwidth}{!}{
    \vspace{-2.5pt}
    \renewcommand{\arraystretch}{1.5}
    \begin{tabular}{l| c| c |c| c| c}
    \hline
    \hline
    \textbf{\normalsize{Estimator}} & \multicolumn{5}{c}{\textbf{\normalsize{Min Entropy}}}\\
    \hline
    \hline
    \bf{} & \bf C58 & \bf C60 & \bf C61 & \bf C62 & \bf C63  \\
    \hline
    \hline
    Most Common Value & 0.99949 & 0.99954 & 0.99949 & 0.99947 & 0.99958 \\
    \hline
    Collision & 0.97489 & 0.97731 & 0.97237 & 0.95915 & 0.97095 \\
    \hline    
    Markov & 0.99986 & 0.99995 & 0.99988 & 0.99964 & 0.99984 \\
    \hline
    Compression & 0.94808 & \bf0.94156 & 0.94419 & 0.96217 & 0.94891 \\
    \hline
    t-Tuple & \bf0.94499 & 0.94499 & \bf0.94219 & \bf0.94358 & \bf0.94643 \\
    \hline    
    LRS & 0.99119 & 0.99895 & 0.97301 & 0.99707 & 0.99845 \\
    \hline
    Multi MCW Prediction & 0.99989 & 0.99978 & 0.99979 & 0.99992 & 0.99958 \\ 
    \hline
    Lag Prediction & 0.99947 & 0.99964 & 0.99961 & 0.99926 & 0.99926 \\
    \hline    
    MultiMMC Prediction & 0.99970 & 0.99984 & 0.99954 & 0.99953 & 0.99941 \\
    \hline
    LZ78Y Prediction & 0.96743 & 0.99983 & 0.99956 & 0.99938 & 0.99948 \\  
    \hline
    \hline 
    \end{tabular}}
    \label{NIST90B_non_IID_test_results}
    \vspace{-10pt}
\end{table}

The NIST SP 800-90B test suite additionally estimates the min-entropy, using two distinct sets of tests, one for TRNGs with IID and one for non-IID sources of entropy. For TRNGs with IID outputs, min-entropy is estimated using the most common value estimate. Here, the most common value is determined and then its proportion in the bit sequence-under-test is computed. The upper bound of the corresponding confidence interval is used as the min-entropy per sample estimate. For binary data, the most common value is either 0 or 1 and the proportion is simply the larger fraction of 0s or 1s in the bit sequence. The values obtained for the five Zynq devices are shown in the first row of Table \ref{NIST90B_non_IID_test_results}, which shows that nearly a full bit of entropy is contained in each output bit.
\begin{table}[t!]
    \caption{NIST SP 800-90B Non-IID and AIS 31 Test Descriptions}  
    \resizebox{\columnwidth}{!}{
    \vspace{-2.5pt}
    \centering 
    \renewcommand{\arraystretch}{1.6}
    \begin{tabular}{l | c }
    \hline
    \hline
    \textbf{\normalsize{Test Name}} & \bf \normalsize{Evaluation Criteria}  \\
    \hline
    \hline
    \bf NIST SP 800-90B \\
    \hline
    \hline  
    Most Common Value & Measures the frequency of the most common value \\
    \hline
    Collision & \makecell{Measures the frequency of repeated values (collisions)} \\
    \hline
    Markov & \makecell{Measures the degree of predictability based on past \\ output values} \\
    \hline
    Compression & Measures the degree of compression possible \\
    \hline
    t-Tuple & \makecell{Evaluates how often sequences of $t$ consecutive \\ symbols repeat} \\
    \hline 
    LRS & \makecell{Detects redundancy by finding the longest \\ repeated substring} \\
    \hline
    MultiMCW Prediction & \makecell{Evaluates predictability by analyzing symbol patterns\\ using multiple "Most Common in Window" predictors} \\
    \hline
    Lag Prediction & \makecell{Analyzes the ability to predict output \\ based on previous outputs at different lags} \\
    \hline  
    MultiMMC Prediction & \makecell{Considers multiple Markov chains and evaluates\\ how well they can predict future values} \\
    \hline
    LZ78Y Prediction & \makecell{Uses the LZ78 compression method to measure \\ predictability of the sequence}\\
    \hline
    \hline
    \bf AIS-31 \\
    \hline
    \hline
    Disjointness & \makecell{Ensures outputs from the entropy source are\\ independent across multiple tests} \\
    \hline
    Monobit & Verifies the balance of 0s and 1s in the bitstream \\
    \hline  
    Poker & \makecell{Analyzes frequency distributions to identify patterns\\ in the data} \\
    \hline
    Runs  & \makecell{Checks the randomness of bit sequences by analyzing\\ runs of consecutive 0s and 1s} \\
    \hline
    Long Run  & \makecell{Identifies overly long runs of consecutive 0s or 1s}  \\
    \hline    
    Autocorrelation & \makecell{Evaluates the dependency between bits \\separated by fixed lags} \\
    \hline
    Uniform Distribution & \makecell{Checks if output values follow a uniform distribution}  \\
    \hline 
    Homogeneity & \makecell{Tests consistency of symbol distributions\\ across subsets of data} \\
    \hline
    Entropy Estimation & \makecell{Measures the unpredictability of the source output} \\
    \hline  
    \hline 
    \end{tabular}}
    \label{90B_non_IID_AIS_Explanation_Table}
\end{table}

We also ran the non-IID test suite to estimate min-entropy, despite the fact that the bit sequences pass the IID tests. The non-IID test suite conservatively estimates min-entropy using a diverse battery of tests, which are listed and briefly explained in the top portion of Table \ref{90B_non_IID_AIS_Explanation_Table}. The smallest min-entropy estimate obtained from one of these tests is used as the estimate for the bit sequence. The results of applying the non-IID test suite to the 10 MByte bit sequences obtained from the five Zynq devices are shown in the remaining rows of Table \ref{NIST90B_non_IID_test_results}, where we have highlighted in bold font the worst-case min-entropy test results. The worst-case values vary between 0.941 to 0.946, which suggests the SiRF TRNG produces a high-quality random bit sequence.

\subsection{AIS 31 Statistical Test Results}
As part of our evaluation process, we applied the AIS 31 test suite as additional validation of the SiRF TRNG. A brief description of the AIS 31 tests is provided in the lower portion of Table \ref{90B_non_IID_AIS_Explanation_Table}.

The results obtained after applying the AIS 31 tests to the 10 MByte bit sequence generated from one of the Zynq devices are shown in Table \ref{AIS31_Results}, which shows that all tests passed. The pass-fail results for the other boards are identical. Similar to the claims made using the NIST SP 800-90B test suite, these results also validate that the SiRF TRNG produces a high-quality random bit sequence.

\begin{table}[t!]
    \caption{Results of the AIS 31 Statistical Test Suite}   
    \resizebox{\columnwidth}{!}{
    \vspace{-2.5pt}
    \centering 
    \renewcommand{\arraystretch}{1.5}
    \begin{tabular}{l c c}
    \hline
    \hline
    \textbf{Test} & \bf Pass rate & \bf Result \\
    \hline
    \hline
    T0 - Disjointness test & 1/1 & Pass \\
    \hline
    T1 - Monobit test & 257/257 & Pass \\
    \hline    
    T2 - Poker test & 257/257 & Pass \\
    \hline
    T3 - Runs test & 257/257 & Pass \\
    \hline
    T4 - Long run test & 257/257 & Pass \\
    \hline    
    T5 - Autocorrelation test & 257/257 & Pass \\
    \hline
    \hline
    \bf Test & \bf Test statistic / Pass condition & \bf Result \\
    \hline
    \hline    
    T6 - Uniform distribution test & \tiny{$|\mathbb{P}(1) - 0.5| = 0.00235/ \boldsymbol{<} \textbf{0.025}$} \vspace{-4pt} & Pass \\ & \tiny{$|\mathbb{P}(01) - \mathbb{P}(11)| = 0.00306 / \boldsymbol{<} \textbf{0.020} $} \\
    \hline 
    T7 - Test for homogeneity & \tiny{$T[0] = 3.329; T[1] = 2.165; / \boldsymbol{<} \textbf{15.13} $} \vspace{-4pt} & Pass \\ \vspace{-4pt} & \tiny{$T[00] = 1.670; T[01] = 3.281;$} \\ & \tiny{$T[10] = 0.077; T[11] = 0.022; / \boldsymbol{<} \textbf{15.13} $} \\ 
    \hline
    T8 - Entropy estimation & \tiny{$H_{1} = 8.00169/ \boldsymbol{\geq} \textbf{7.976} $} & Pass \\      
    \hline  
    \hline 
    \end{tabular}}
    \label{AIS31_Results}
    \vspace{2pt}
\end{table}

\subsection{DieHarder Statistical Test Results}
For completeness, we generate random bit sequences from the five Zynq board and use them as input to the DieHarder test suite \cite{DieHarder}. The DieHarder tests are applied by redirecting the SiRF PUF-TRNG bit sequences directly to the DieHarder test tool. Given the bit generation rate is 2.67 Mbits per second, the five Zynq devices used in the experiments ran for more than 20 days before enough bits were generated to run all 116 DieHarder tests. No test failures occurred for any test and for any of the five bit sequences generated by the Zynq devices, and only 18 'WEAK' results were observed. For comparison, we also subjected the random bit sequences generated by the OpenSSL pseudo-random number generator to the DieHarder test suite in five separate experiments and observed no failures, and only 19 'WEAK' results. Therefore, the quality of the SiRF TRNG bit sequences is similar to those produced by OpenSSL.


\subsection{Statistical Analysis of Dynamic Entropy}
As indicated earlier, the nonces used to randomize parameters to the 64-bit LFSR Challenge generator and Sponge Function modules represent the dynamic entropy component of the TRNG. We collected 10 sequences of 100,000 bits from the five devices and ran the NIST SP 800-22 test suite on the sequences. All applicable NIST statistical tests passed, as well as the p-value-of-the-p-value tests. Therefore, the distillation operation, which XORs 12 consecutive low-order bits of the path delay measurements, is capable of generating bit sequences of high statistical quality.

\subsection{Analysis of RC and TCC SiRF PUF-TRNG Parameters}
\begin{figure}[t!]
    \centering    \includegraphics[width=3.4in,keepaspectratio=true]{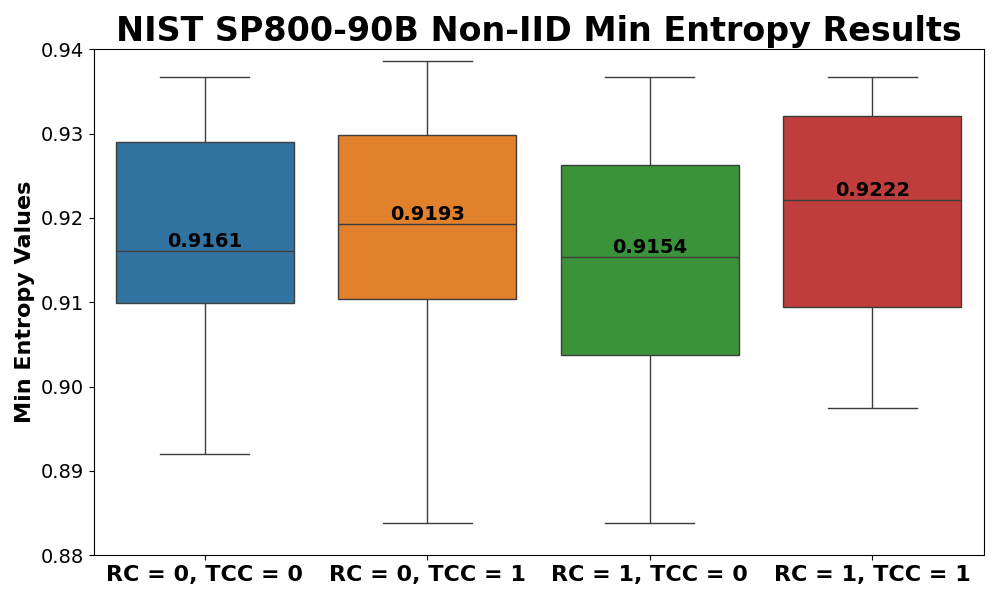}
     \caption{Box plots showing the NIST SP 800-90B Non-IID min-entropy values from 25 boards. The horizontal axis represents the different configurations of Range Constant (RC) and Trim Code Constant (TCC), indicating whether each is turned ON or OFF. The median value for each configuration is displayed above the corresponding boxplot.}
     \label{RC_TCC_Analysis}
     \vspace{-5pt}
\end{figure}

We carry out a special set of tests that evaluate the benefit of randomizing two parameters used by the Sponge Function loop shown in Fig. \ref{Figure:SiRF_TRNG_algorithm}, namely, the Range Constant (RC) in the GPEV module and the Trim Code Constant (TCC) in the SF module. The DV used in these evaluations are held constant across the experiments while the RC and TCC parameters are either randomly varied, e.g., RC = 1 or held constant, e.g., RC = 0, as shown by the labels along the bottom of the box plot graph shown in Fig. \ref{RC_TCC_Analysis}. The box plot characterizes the min-entropy results from the NIST SP 800-90B test suite for bit sequences generated by the SiRF PUF-TRNG algorithm using DV collected from 25 devices. The best result is obtained when both RC and TCC are randomly varied, as exemplified by the larger median value for the right-most box plot when compared with the others. From the results, it is clear that the improvement is marginal, suggesting that most of the min-entropy is obtained by the SF Chaining operation. Despite the small improvement, randomizing these parameters adds uncertainty to the processing operations within the Sponge Function loop, and therefore, improves attack resilience. 

\subsection{Comparison with other TRNGs}
In this section, we compare the statistical test results of the SiRF PUF-TRNG with two stand-alone TRNGs proposed in \cite{Ingrid2022} and \cite{Chaos2020}. These particular TRNGs are referenced because the authors apply the NIST SP 800-90B and AIS 31 test suites to their bit sequences. We were not able to find any unified PUF-TRNG architecture papers that provided a comprehensive evaluation using these test suites. The comparison in Table \ref{Comparison Table} shows that the SiRF PUF-TRNG has a smaller bit generation rate, but compares favorably in terms of the other metrics. The larger footprint associated with the SiRF PUF-TRNG is due to the inclusion of the PUF security function. 

\begin{table}[t!]
    \caption{Comparison with Other FPGA TRNG Designs}   
    \vspace{-2.5pt}
    \centering
    \renewcommand{\arraystretch}{1.5}
    \begin{tabular}{c c c c c c}
    \hline
    \hline
    \vspace{-10pt} \\
    \textbf{Design} & {\bf Device} & \makecell{\bf Entropy\\ \bf Est. } & \bf Area & \makecell{\bf ThrPut \\ \bf [Mbps]} & \makecell{\bf Power \\ \bf [mW]} \\
    \vspace{-10pt} \\
    \hline
    \hline 
    \textbf{SiRF} & \makecell{Zynq \\ 7010} & $ \geq 0.999$ & \makecell{5842 LUTs \\ 4377 FFs \\ 32 CARRY4s} & 2.5 & 27 \\
    \hline
    \makecell{TROT \\ \cite{Ingrid2022}} & \makecell{Zynq \\ 7000} & $ \geq 0.999$ & \makecell{32 LUTs \\ 55 FFs \\ 17 CARRY4s} & 12.5 & 9.5 \\
    \hline
    \makecell{CHAOS\\ \cite{Chaos2020}} & Virtex-6 & $ \geq 0.983$ & \makecell{53 LUTs \\ 22 FFs} & 1600 & 2.05 \\
    \hline
    \hline 
    \end{tabular}
    \label{Comparison Table}
    \vspace{2pt}
\end{table}

\subsection{SiRF PUF-TRNG Bit Generation Rate and Resource Utilization}

The path timing operation carried out in Phase 1 (from Fig. \ref{Figure:SiRF_TRNG_algorithm}) takes approximately 50 milliseconds using a 50 MHz clock frequency, while the linear operations carried out in Phase 2 by the four post-processing modules are able to execute in approximately 600 microseconds per iteration, yielding a bit generation rate of approximately 2.67 Mbits per second. Note that increasing the clock frequency would increase the bit generation rate proportionally, e.g., 100 MHz would double the rate.

The SiRF PUF-TRNG implementation on the Zynq FPGAs utilizes 5,842 LUTs and 4,377 FFs, and requires two DSP primitives to implement multiplication in two modules of the PUF-TRNG algorithm. Resource utilization is nearly identical (within 5\%) to the utilization required for the SiRF PUF stand-alone. The BRAM utilization is 24 KBytes, which is 4,096 bytes larger than that required by SiRF PUF stand-alone. The additional 4,096 bytes are needed to accommodate the SF Chaining operation.

\section{Conclusions} \label{Section:Conclusions}
In this paper, we present a unified SiRF PUF-TRNG architecture that utilizes static entropy from a strong PUF and dynamic entropy derived from path delay noise. The novel inclusion of a soft-data sponge function enhances randomness and efficiency, while resilience against temperature-voltage attacks is achieved through the GPEV post-processing module. Extensive evaluation using the NIST SP 800-22, NIST SP 800-90B, AIS 31, and DieHarder test suites demonstrates the TRNG's robust statistical performance and high min-entropy. The architecture demonstrates a compact and resource-efficient approach, reusing over 95\% of the SiRF PUF's components, with a bit generation rate of 2.67 Mbps and minimal resource overhead. The proposed approach advances the integration of PUF and TRNG capabilities, providing a reliable and scalable solution for secure hardware systems.


\ifCLASSOPTIONcaptionsoff
  \newpage
\fi

\bibliographystyle{IEEEtran}
\bibliography{IEEEabrv,references}

\vspace{-40pt}

\begin{IEEEbiography}[{\includegraphics[width=0.63in,height=1.0in,clip,keepaspectratio]{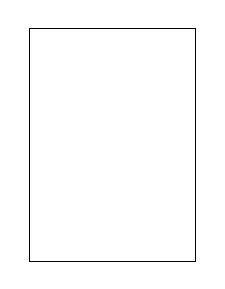}}]{Rachel Cazzola}
is pursuing a Ph.D. in Computer Engineering with an emphasis on Hardware Security in the Electrical and Computer Engineering department at the University of New Mexico.
\end{IEEEbiography}

\vspace{-40pt}

\begin{IEEEbiography}[{\includegraphics[width=0.63in,height=1.0in,clip,keepaspectratio]{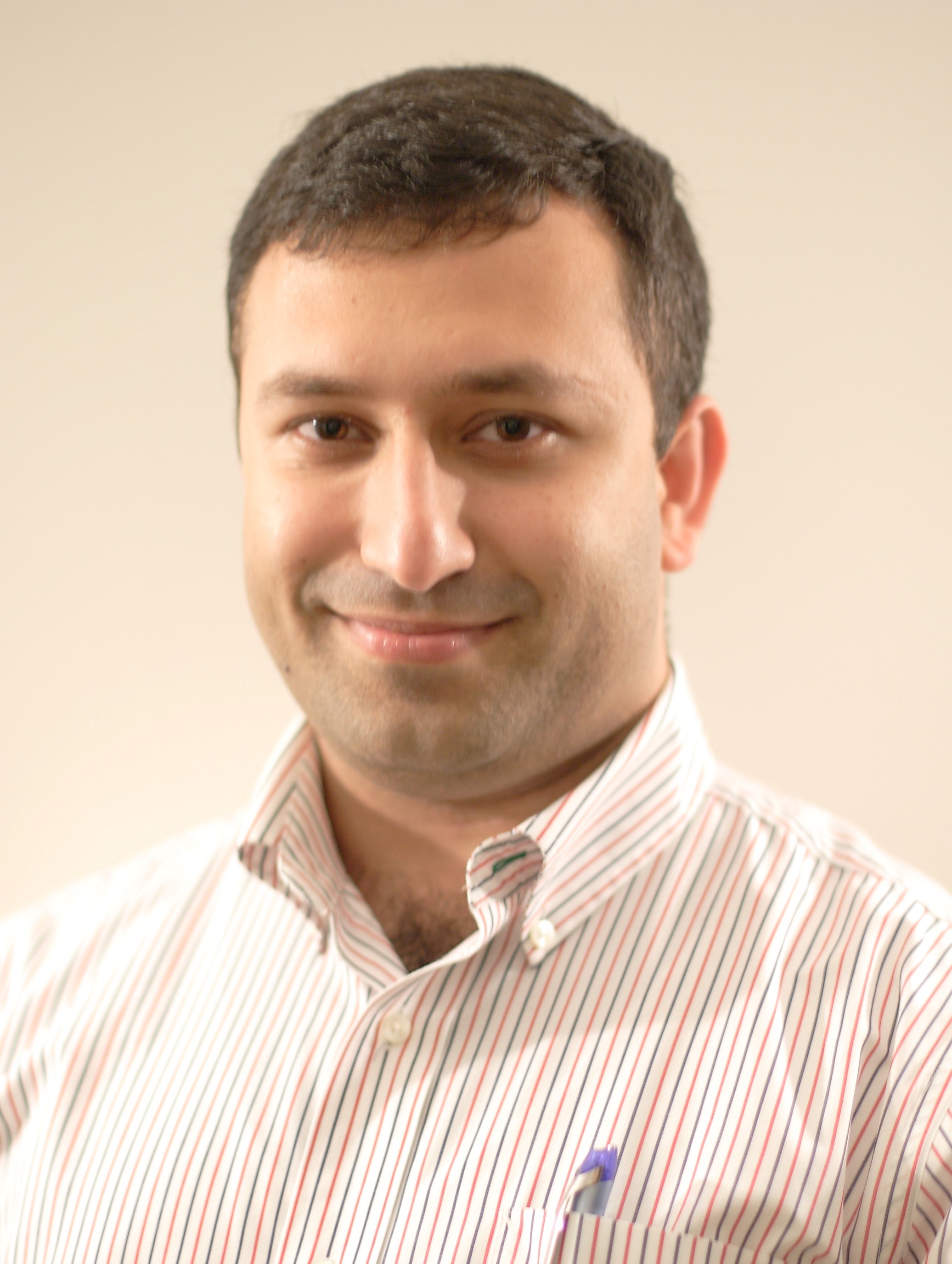}}]{Cyrus Minwalla}
is a Principal Researcher at the Bank of Canada. He received his M.S. and Ph.D. degrees in Computer Engineering from York University in Canada. Cyrus was awarded the Bank of Canada’s Award of Excellence in 2020 and 2022, and was selected as NRC’s Top Scientist Under 40 in 2017. His research interests include digital currencies, cryptography, embedded devices, and Internet-of-Things. 
\end{IEEEbiography}

\vspace{-40pt}

\begin{IEEEbiography}[{\includegraphics[width=0.63in,height=1.0in,clip,keepaspectratio]{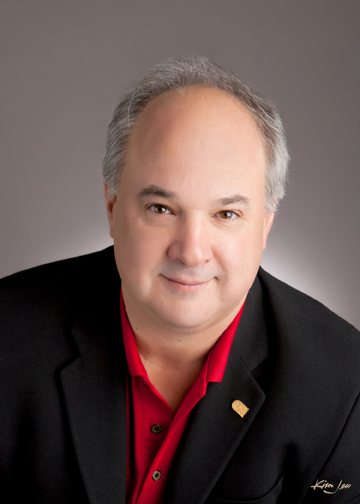}}]{Jim Plusquellic}
is a Professor in Electrical and Computer Engineering at the University of New Mexico. He received both his M.S. and Ph.D. degrees in Computer Science from the University of Pittsburgh. Professor Plusquellic received an "Outstanding Contribution Award" from IEEE Computer Society in 2012 and 2017 for co-founding and for his contributions to the Symposium on Hardware-Oriented Security and Trust (HOST).
\end{IEEEbiography}

\end{document}